\documentclass[12pt]{article}

\usepackage{amsmath}
\usepackage{amssymb}
\usepackage{amsfonts}
\usepackage{lscape}

\begin{document}

\fontsize{12}{6mm}\selectfont
\setlength{\baselineskip}{2em}

$~$\\[.35in]
\newcommand{\dss}{\displaystyle}
\newcommand{\raro}{\rightarrow}
\newcommand{\be}{\begin{equation}}

\def\sech{\mbox{\rm sech}}
\def\sn{\mbox{\rm sn}}
\def\dn{\mbox{\rm dn}}
\thispagestyle{empty}

\begin{center}
{\Large\bf Geometric Properties of  } \\    [2mm]
{\Large\bf Quantum Phases } \\    [2mm]
\end{center}

\vspace{1cm}
\begin{center}
{\bf Paul Bracken}                        \\
{\bf Department of Mathematics,} \\
{\bf University of Texas,} \\
{\bf Edinburg, TX  }  \\
{78541-2999}
\end{center}

\vspace{3cm}
\begin{abstract}
The Aharonov-Anandan phase is introduced from a physical
point of view. Without reference to any dynamical equation,
this phase is formulated by defining an appropriate connection
on a specific fibre bundle. The holonomy element gives
the phase. By introducing another connection, the
Pancharatnam phase formula is derived following a different
procedure. 
\end{abstract}

\vspace{2mm}

\vspace{2mm}

\vspace{2mm}

\newpage

The discovery and subsequent interest in the Berry
phase {\bf [1]} has been relatively recent with respect to
the actual period over which quantum mechanics has been in use.
Beyond its physical significance, it has generated a
great deal of interest into more geometric approaches
to quantum mechanics as well as applications of many ideas
from the area of differential geometry, in particular,
fibre bundles and connections. General relativity
and Yang-Mills gauge theories are also examples in
which geometrical techniques enter into the study of
these theories directly. 
In fact, quantum mechanics can be looked at geometrically.
Here ${\cal H}$ will refer to a Hilbert space in
general and any quantum system carries the structure of a
K\"{a}hler manifold. Even so, the space ${\cal H}$ is
not the quantum analog of a classical phase space.
In what follows, elements of ${\cal H}$ will be denoted by $\psi$
or $| \psi \rangle$, but the bracket will always appear when 
the inner product is invoked.
The Berry phase is known to 
depend on the geometric structure of the parameter space itself,
so the phase is really a geometric property.
The purpose here is to further explore the phase by
looking for ways of formulating the ideas in a more
intrinsic manner, and to present a different development of 
the integral formula for the Pancharatnam phase.
Simon {\bf [2]} interpreted this phase as the holonomy
of the adiabatic connection in the bundle appropriate to
the evolution of the adiabatic eigenstate and expressed 
it as an integral over a connection one-form. 
Aharonov and Anandan {\bf [3]} defined a geometric
phase during any cyclic evolution of a quantum system
which depends only on the topological features and the
curvature of the quantum state space. 
Wilczek and Zee have considered a nonabelian extension of the phase
{\bf [4]}. Considered in
this way, the origin of the geometric phase is 
due to the parallel transport of a state vector on the
curved surface. This is a fundamental notion in modern
differential geometry since it is directly related
to the concept of a connection, and there are often
several ways in which a connection may be defined {\bf [5]}.
Some work which is related to the results here has
been done by making use of geodesics {\bf [6]},
a related but different approach from this one.

To see how the idea of a connection can arise physically 
in this context and to
define a connection from a physical point
of view, suppose a state vector $| \psi \rangle$
is an element of a Hilbert space ${\cal H}$ 
which evolves according to the Schr\"odinger
equation
$$
i \frac{\partial}{\partial t} |\psi (t) \rangle
= {\bf H} (t) | \psi (t) \rangle,
\eqno(1)
$$
where ${\bf H} (t)$ is a linear operator,
which need not be Hermitian. A new state vector
$|\phi (t) \rangle$ can be defined which has
a dynamical phase factor removed. It is given by
$$
| \phi (t) \rangle = \exp ( i \int_0^t h (\tau ) )
| \psi (t) \rangle,
\eqno(2)
$$
such that $h(t)$ is defined as the real quantity
$$
h (t) = Re \langle \psi (t) | {\bf H} (t) |
\psi (t) \rangle.
\eqno(3)
$$
Differentiating $| \phi (t) \rangle$
with respect to $t$ and requiring that
$|\phi (t) \rangle$ satisfies the
Schr\"odinger equation,
$$
i \frac{\partial}{\partial t} |\phi (t) \rangle =
- h (t) | \phi (t) \rangle
+ i \exp (i \int_0^t h ( \tau ) \, d \tau )
\frac{\partial}{\partial t} | \psi (t) \rangle
= - h (t) |\phi (t) \rangle + H (t) | \phi (t) \rangle.
$$
Therefore, $|\phi (t) \rangle$ satisfies the
equation
$$
\frac{\partial}{\partial t} | \phi (t) \rangle =
i [ {\bf H} (t) - h(t) ] |\phi (t) \rangle.
\eqno(4)
$$
Since $h (t)$ is the real part of $\langle \psi |
{\bf H} | \psi \rangle$, upon contracting with
$\langle \phi (t) |$, the right-hand side 
must be real hence
$$
Im \, \langle \phi (t) | \frac{\partial}{\partial t} 
| \phi (t) \rangle = 0.
\eqno(5)
$$
This can be regarded as a parallel transport rule.

To approach this in a more intrinsic manner,
the idea of a fibre bundle will be introduced {\bf [6]}.
Any two vectors $\psi$, $\phi$ in ${\cal H}$
such that $\psi = c \phi$ where $c \in \mathbb C$
are physically equivalent since they define the
same state and we write $\psi \sim \phi$.
Thus, the correct phase space of a quantum system
is the space of rays in the space ${\cal H}$
and this is denoted by writing ${\cal P} ( 
{\cal H}) = {\cal H} / \sim$. The notation
$\sim$ denotes the elements of ${\cal H}$
which differ only by a phase and is 
referred to as projective Hilbert space here.
A canonical projection operator $\Pi$
can then be defined between these two spaces as
$$
\Pi : {\cal H} \rightarrow {\cal P} ( {\cal H}),
\eqno(6)
$$
An element of the space ${\cal P} ( {\cal H})$
may be denoted by $[\psi] = \Pi (\psi)$. Thus $\Pi$
maps $\psi$ to the ray on which it lies.
The fibres $\Pi^{-1} ([ \psi])$ are one-dimensional,
and this type of vector bundle is referred to
as a complex line bundle. The unit sphere is 
a subset of ${\cal H}$ and is given by
$$
S ({\cal H}) = \{ \psi \in {\cal H} 
| \langle \psi | \psi \rangle = 1 \} \subset {\cal H}.
\eqno(7)
$$
Thus, we can write equivalently ${\cal P} ( {\cal H})
= S ( {\cal H}) / \sim$. Suppose $| \phi (s) \rangle$ 
is a curve in ${\cal H}$ and define
$$
| m \rangle = \frac{d}{ds} | \phi (s) \rangle, 
\eqno(8)
$$
which denotes the tangent vector to this curve.
In terms of $|\phi (s) \rangle$ and $|m \rangle$,
we can define
$$
A_s = \frac{Im \, \langle \phi | m \rangle}{\langle 
\phi | \phi \rangle}. 
\eqno(9)
$$
A transformation acting on $| \phi (s) \rangle$
which has the form $|\phi (s) \rangle 
\rightarrow | \hat{\phi} (s) \rangle = \exp ( i \alpha (s)) | \phi (s) \rangle$
has the structure of a gauge transformation. 
Differentiating the transformed $| \phi (s) \rangle$,
with respect to $s$ gives
$$
| \hat{m} \rangle = e^{i \alpha} \frac{d}{ds} | \phi (s) \rangle
+ i \frac{d \alpha}{ds} e^{i \alpha} | \phi (s) \rangle .
\eqno(10)
$$
From (10), the transformed function $\hat{A}_s$ can be
obtained in the form
$$
\hat{A}_s = \frac{Im \, \langle \hat{\phi} | \hat{m} \rangle}
{\langle \hat{\phi} | \hat{\phi} \rangle} = A_s + i \frac{d \alpha}{ds}.
\eqno(11)
$$
Therefore, $A_s$ which is defined by expression (9)
transforms like the vector potential in electrodynamics.
The parallel transport law (5) then states that
$A_s$ vanishes along the actual curve $|\phi (s) \rangle$
which is taken by the quantum system in the quantum space.

Let $| \psi (t) \rangle$ be a solution of the
Schr\"odinger equation which is cyclic. This
means that it returns to the initial ray after
a given time $\tau$ and as well specifies a curve in
${\cal H}$. Under the map $\Pi$
this curve is mapped to a closed curve in ${\cal P} (
{\cal H})$. Given a closed curve $\sigma (s)$ in
${\cal P} ( {\cal H})$, let us consider the
curve in ${\cal H}$, 
which is traced out by the state vector $| \phi (s) \rangle$.
Using the parallel transport law, the curve
is determined by the condition that $A_s =0$ holds
along the actual curve. Define the integral
$$
\gamma = \oint_{\Gamma} A_s \, ds.
\eqno(12)
$$
The path $\Gamma$ in (12) is traced out along the curve
$|\phi (s) \rangle$ in the space ${\cal H}$ which has
been made closed by the vertical curve joining
$| \phi ( \tau ) \rangle$ to $| \phi (0) \rangle$.
The segment along $| \phi (s) \rangle$ generates
the actual evolution of the system, but by the parallel
transport law (5), it is clear that $A_s = 0$
along this segment. 
It is left to the vertical segment of the trajectory
to contribute the phase difference between the
states $| \phi (0) \rangle$ and $| \phi ( \tau ) \rangle$.
The integral in (12) is gauge invariant on account
of the transformation rule (11), and it can therefore
be considered an integral on ${\cal P} ( {\cal H})$.
By Stokes theorem, $\gamma$ can also be expressed in the form
$$
\gamma = \int_S d A_s = \int_S F,
\eqno(13)
$$
such that $S$ is any surface in ${\cal P} ( {\cal H})$
bounded by the closed curve $\sigma (s)$ in ${\cal P}
( {\cal H})$. The field strength $F$ which appears
in (13) is a gauge invariant two-form as well.
From this, it can be seen that $\gamma$ in (12)
and (13) is a geometrical quantity depending 
on the geometric curve $\sigma (s)$. This is the version
of Berry's phase in a cyclic evolution of the 
quantum system.

Let us formulate this in a more geometric way
by looking for an appropriate connection in a
principle $U (1)$-bundle, $S ( {\cal H})
\rightarrow {\cal P} ( {\cal H})$.
To introduce a connection we have to define
a subspace of horizontal vectors.
Identifying the tangent space $T_{\psi}
S ( {\cal H})$ as a linear subspace in
${\cal H}$, a decomposition exists of the form
$$
T_{\psi} S ( {\cal H}) = V_{\psi} 
+ H_{\psi}.
\eqno(14)
$$
Hence the subspaces of vertical and horizontal
vectors are linear subspaces of ${\cal H}$.
A fibre $\Pi^{-1} (\psi)$ consists of all
vectors of the form $e^{i \lambda} \psi$.
The vertical subspace $V_{\psi}$ in (14) can then be 
defined by
$$
V_{\psi} = \{ i \lambda \psi | \, \lambda \in \mathbb R \},
$$
which can be identified with $u (1)$. To define a
natural connection, let $X$ be a vector tangent to
$S ( {\cal H})$ at $\psi$. Then $X$ is called a
horizontal vector with respect to a natural
connection if
$$
\langle \psi | X \rangle = 0.
$$
The set of horizontal vectors at $\psi$ in (14) can be
defined as follows
$$
H_{\psi} = \{ X \in {\cal H} | \langle  \psi | X \rangle = 0 \}.
$$
A curve $t \rightarrow \psi (t) \in S ( {\cal H})$ is
horizontal if
$$
\langle \psi (t) | \dot{\psi} ( t) \rangle = 0.
$$
Since $\langle \psi | \psi \rangle =1$, differentiating
with respect to the parameter, $\langle \dot{\psi} | 
\psi \rangle + \langle \psi | \dot{\psi} \rangle =0$
which implies that $Re \, \langle \psi (t) | \dot{\psi} (t) \rangle =0$
and so the horizontal condition can be expressed as
$$
Im \, \langle \psi (t) | \dot{\psi} (t) \rangle =0.
$$
A connection one-form ${\cal A}$ in a principal $U(1)$
bundle $S ( {\cal H}) \rightarrow {\cal P} ({\cal H})$
is a $u (1)$-valued one-form on $S ({\cal H})$.
Take an element $X \in S ( {\cal H}) \subset {\cal H}$
and define in $u(1)$
$$
A_{\psi} (X) = i \, Im \, \langle \psi | X \rangle.
$$
Therefore, $X$ is horizontal at a point $\psi \in 
S ( {\cal H})$ if ${\cal A}_{\psi} (X) = 0$.
Consider now a local connection form $A$ on
${\cal P} ( {\cal H})$ such that $\Psi :
{\cal P} ( {\cal H}) \rightarrow  S ( {\cal H})$ is
a local section. The pull back of ${\cal A}$
$$
A = i \Psi^* {\cal A},
$$
defines a local connection one-form on ${\cal P} ( {\cal H})$.
This implies the local connection $A$ in gauge $\psi$ 
can be written
$$
A = i \langle \psi | d \psi \rangle.
\eqno(16)
$$
Once the connection has been defined as in (16),
the corresponding holonomy element may be
computed from $A$ as
$$
\Phi ( C ) = \exp ( i \oint_C A ),
\eqno(17)
$$
where $C$ is a closed curve in ${\cal P} ( {\cal H})$.

Some additional information will be needed to show the 
remaining result. Take two nonorthogonal vectors
$\psi_1$, $\psi_2 \in S ( {\cal H})$.
The phase of their scalar product will be called
the {\em relative} phase or phase difference
between $\psi_1$ and $\psi_2$. Thus
$\langle \psi_1 | \psi_2 \rangle = r e^{i \alpha_{12}}$
so $\alpha_{12}$ is the phase difference between
$\psi_1$ and $\psi_2$. Hence $\psi_1$ and $\psi_2$
are in phase or parallel if $\langle \psi_1 | \psi_2 \rangle$
is real and positive. There is a relation then between
any two nonorthogonal vectors $\psi \sim \phi$ if and only if
they are in phase. This procedure is yet another way
of equipping a principal $U (1)$ fibre bundle
$S ( {\cal H}) \rightarrow {\cal P} ( {\cal H})$ with a
connection.

Furthermore, if $p_1$ and $p_2$ are two
points in ${\cal P} ( {\cal H})$, then let
$\psi_1$ and $\psi_2$ be two arbitrary nonorthogonal
state vectors in $S ( {\cal H})$ projecting down to
$p_1$ and $p_2$, respectively. A real plane in ${\cal H}$
can be defined by the pair $\psi_1$ and $\psi_2$
in the following way
$\{ \psi = \xi_1 \psi_1 + \xi_2 \psi_2 | \xi_1, \xi_2 \in \mathbb R \}
\subset {\cal H}$.
This gives a natural way to obtain a geodesic since the
intersection of any real plane with the unit sphere
$S ( {\cal H})$ is a great circle. This defines a geodesic
on $S ( {\cal H})$ with respect to the metric induced from
${\cal H}$. A geodesic on the sphere $S ( {\cal H})$
projects to a geodesic on ${\cal P}( {\cal H})$,
and hence each geodesic on ${\cal P} ({\cal H})$
is a closed curve since it is the projection of a
closed curve. Thus, a geodesic joining $\psi_1$ and
$\psi_2$ on $S ( {\cal H})$ is an arc of a great
circle passing through $\psi_1$ and $\psi_2$
and is parametrized by an angle $\theta \in
[0, 2 \pi)$ such that
$$
\psi ( \theta ) = \xi_1 (\theta) \psi_1 
+ \xi_2 ( \theta ) \psi_2.
\eqno(18)
$$
Now define the real parameter $a = Re \, \langle \psi_1 | \psi_2 \rangle$
and suppose that $a >0$. The normalization condition
$\langle \psi ( \theta) | \psi ( \theta ) \rangle =1$
takes the form
$$
\xi_1^2 + 2a \xi_1 \xi_2 + \xi_2^2 -1 =0.
\eqno(19)
$$
In terms of the angle $\theta$, the coefficients 
$\xi_1$ and $\xi_2$ can be written as
$$
\xi_1 ( \theta ) = \cos \theta - \frac{a}{\sqrt{1- a^2}} \sin \theta,
\qquad
\xi_2 ( \theta ) = \frac{1}{\sqrt{1 - a^2}} \sin \theta,
\eqno(20)
$$
which satisfy (19).
Moreover, $\psi (0) = \psi_1$ and $\psi (\theta_0 ) = \psi_2$,
where the angle $\theta_0$ is defined by $\cos \theta_0 = a$
such that $\theta_0 \in [0, \pi /2 )$.

It is remarkable that the Pancharatnam phase 
can be expressed as a line integral of $A_s$ with
the use of the geodesic rule.
Let $| \phi_1 \rangle$ and $| \phi_2 \rangle$ be any
two nonorthogonal states in ${\cal H}$ with phase
difference $\beta$.
Let $| \phi (s) \rangle$ be any geodesic
curve connecting $| \phi_1 \rangle$ to $| \phi_2 \rangle$
so that $| \phi (0) \rangle = | \phi_1 \rangle$
and $| \phi (1) \rangle = |\phi_2 \rangle$.
Then the phase difference $\beta$ is given by
$$
\beta = \int A_{s} \, ds,
\eqno(21)
$$
where $A_s$ is given by the natural connection (9).

Consider two points $p_1$, $p_2 \in {\cal P} ( {\cal H})$,
and let $\sigma$ be the shorter arc of the geodesic
which connects $p_1$ and $p_2$. Suppose $\tilde{\sigma}
: t \rightarrow \psi (t) \in S ( {\cal H})$ is a
horizontal lift of $\sigma$ with respect to the
natural connection in the principal fibre bundle
$S ( {\cal H}) \rightarrow {\cal P} ( {\cal H})$.
Then a parallel transport of $\psi$ keeps
$\psi (t)$ in phase with $\psi (0)$.
To see this, let $C$ be a geodesic in $S ( {\cal H})$
projecting to $\sigma$ in ${\cal P} ( {\cal H})$.
Any geodesic on the unit sphere $S ({\cal H})$
is uniquely defined by a real plane in ${\cal H}$
spanned by two vectors $\psi_1$ and $\psi_2$.
The shorter arc of the closed geodesic can be
written as in (18) and is a horizontal lift of
$\sigma$ if and only if $\langle \psi_1 | \psi_2 \rangle$ 
is real and positive. Thus, $\psi_1$ and $\psi_2$ are in 
phase. Using (18), we can work out $\langle 
\psi ( \theta_1 ) | \psi ( \theta_2 ) \rangle$
with $\langle \psi_1 | \psi_2 \rangle =a$ and this is
$$
\langle \psi (\theta_1 ) | \psi ( \theta_2 ) \rangle
= \xi_1 ( \theta_1 ) \xi_1 ( \theta_2 )
+ \xi_2 ( \theta_1 ) \xi_1 ( \theta_2) 
\langle \psi_2 | \psi_1 \rangle 
+ \xi_1 ( \theta_1) \xi_2 (\theta_2) 
\langle \psi_1 | \psi_2 \rangle + 
\xi_2 ( \theta_1 ) \xi_2 ( \theta_2)
$$
$$
= \cos \theta_1 \, \cos \theta_2 
+ \frac{a^2}{1 - a^2} \sin \theta_1 \, \sin \theta_2
+ \frac{1}{1 - a^2} \sin \theta_1 \, \sin \theta_2
$$
$$
- \frac{a^2}{1-a^2} \sin \theta_1 \, \sin \theta_2 
- \frac{a^2}{1 - a^2} \sin \theta_1 \, \sin \theta_2
$$
$$
= \cos ( \theta_1 - \theta_2 ) > 0,
$$
since $\theta_1, \theta_2 \in [ 0, \theta_0 ]$
and $\theta_0$ is given by solving $\cos \theta_0 =a$.
Therefore, any two points belonging to the horizontal
lift $\tilde{\sigma}$ are in phase.

To finish the proof of (21), carry 
out a gauge transformation $| \phi (s) \rangle
= \exp ( i \alpha (s)) | \tilde{\phi} (s) \rangle$
of the horizontal lift $| \tilde{\phi} (s) \rangle$ of
the geodesic in ${\cal P}$.
where $\alpha (s)$ is chosen such that $\alpha (0) =0$ 
and $\alpha (1) = \beta$. Then $| \phi (s) \rangle$
remains a geodesic curve, since the geodesic
equation is gauge covariant and connects $| \phi_1 \rangle$
to $| \phi_2 \rangle$.
Thus since $\tilde{A}_s$ is zero on the horizontal
curve, the  right-hand side of (21)
can be integrated to give 
$$
\int_0^1 \frac{d \alpha (s)}{ds} \, ds
= \alpha (1) - \alpha (0) = \beta.
$$
This completes the proof.

\newpage
\begin{center}
{\bf References}
\end{center}
\noindent
$[1]$ M. V. Berry, Proc. R. Soc. A 392 (1984) 45.  \\
$[2]$ B. Simon, Phys. Rev. Lett. {\bf 51} (1983) 2167.   \\
$[3]$ Y. Aharonov and J. Anandan, Phys. Rev. Lett.
{\bf 58}, (1987) 1593, Phys. Rev. Lett. {\bf 65} (1990) 1697. \\
$[4]$ F. Wilczek and A. Zee, Phys. Rev. Lett. {\bf 52} (1984) 2111.  \\
$[5]$ I. J. R. Aitchison and K. Wanelik, Proc. R. Soc. {\bf 439}, 
(1992) 25.   \\
$[6]$ A. K. Pati, Phys. Lett. {\bf 202}, (1995) 40.   \\
$[7]$ D. Chru\'sci\'nski and A. Jamiolkowski,
Geometric Phases in Classical and Quantum Mechanics, Birkh\"auser, (2004).  \\

\end{document}